\documentclass[letterpaper]{article} % DO NOT CHANGE THIS
\usepackage{aaai2027}
\nocopyright
\usepackage[hyphens]{url} % DO NOT CHANGE THIS
\usepackage{graphicx} % DO NOT CHANGE THIS
\usepackage{natbib} % DO NOT CHANGE THIS AND DO NOT ADD ANY OPTIONS TO IT
\usepackage{caption} % DO NOT CHANGE THIS AND DO NOT ADD ANY OPTIONS TO IT
\usepackage{microtype}
\usepackage{amsmath,amssymb,amsthm,mathtools}
\usepackage{enumitem}
\usepackage{multirow}
\usepackage{subcaption}
\usepackage{booktabs}
\usepackage{arydshln}
\usepackage{xcolor}
\usepackage{algorithm}
\usepackage{algpseudocode}
\usepackage[most]{tcolorbox}
\usepackage{framed}
\usepackage{soul}
\usepackage{colortbl}

\definecolor{lightgray}{rgb}{0.88,0.92,0.98}
\definecolor{defred}{rgb}{0.88,0.2510,0.3294}
\definecolor{defblue}{RGB}{0,113,188}

\newtcolorbox{PromptBox}[1]{
    enhanced,
    title={#1},
    colframe=black!75!white,
    colback=gray!3!white,
    colbacktitle=black!75!white,
    coltitle=white,
    fonttitle=\bfseries\sffamily\large,
    fontupper=\rmfamily,
    boxrule=0.8pt,
    arc=3mm,
    left=4mm, right=4mm, top=3mm, bottom=3mm,
    toptitle=2mm, bottomtitle=2mm,
    breakable
}

\title{EmAvatar: Multimodal Empathetic Response Generation via Conflict Resolution and Expressive Guidance}
\author{
    Xiaolin Chen\textsuperscript{\rm 1},
    Xuemeng Song\textsuperscript{\rm 2},
    Jinlan Fu\textsuperscript{\rm 3},
    Weili Guan\textsuperscript{\rm 4},
    Mong-Li Lee\textsuperscript{\rm 1},
    Wynne Hsu\textsuperscript{\rm 1}
}
\affiliations{
    \textsuperscript{\rm 1}National University of Singapore\\
    \textsuperscript{\rm 2}Southern University of Science and Technology\\
    \textsuperscript{\rm 3}Fudan University\\
    \textsuperscript{\rm 4}Harbin Institute of Technology (Shenzhen)\\
}

\begin{document}
\maketitle

\begin{abstract}
Avatar-based multimodal empathetic response generation has emerged as a pivotal capability in human-centric systems, aiming to recognize user emotions and synthesize responses with synchronized text, audio, and talking-face video. 
Despite recent progress, existing methods still suffer from three critical limitations: (1) overlooking conflicting emotions across modalities, (2) lacking explicit multimodal synthesis guidance, and (3) neglecting inherent error propagation of multimodal response generation. 
To address these limitations, we propose EmAvatar, a novel framework for precise emotion perception and expressive response generation. 
It first performs deliberative multimodal emotion recognition by exposing inter-modal prediction conflicts and then initiates a multi-round QA process between a Conflict Inspector and an Evidence Collector to gather evidence for conflict resolution, leading to a robust, evidence-aware prediction.
Regarding response generation, EmAvatar first synthesizes a composite script that couples the textual response with an expressive instruction. Moreover, to ensure high-quality synthesis, an iterative refinement mechanism evaluates and revises the script until it aligns with predefined criteria, serving as reliable guidance for subsequent audio and video synthesis.
Extensive experiments across four tasks demonstrate that EmAvatar outperforms state-of-the-art methods. Our code will be publicly released.
\end{abstract}

\section{Introduction}
\begin{figure}[!t]
    \centering
    \includegraphics[scale=0.6]{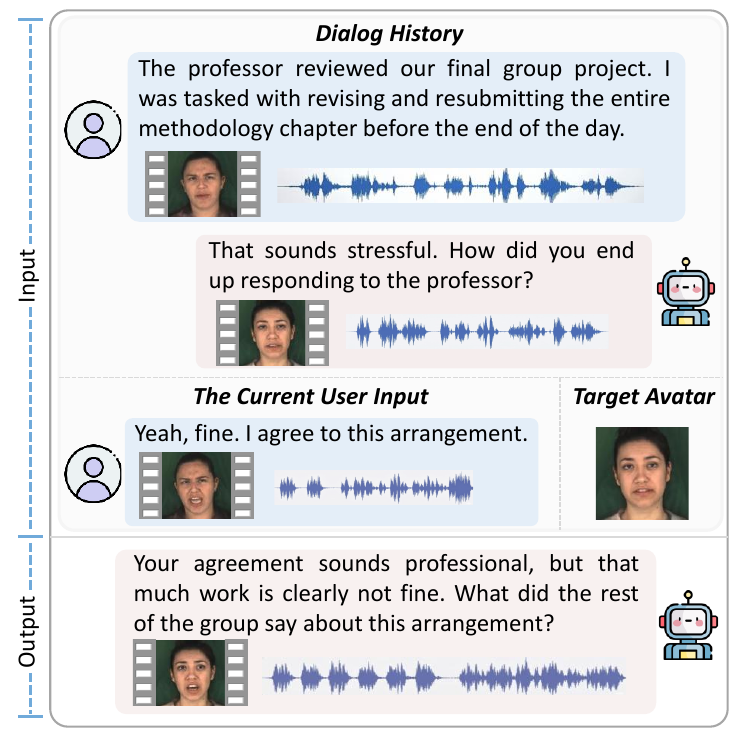}
    \caption{Illustration of avatar-based multimodal empathetic response generation.}
    % \vspace{-1em}
    \label{Model_task}
\end{figure}

As conversational systems are increasingly deployed in emotionally sensitive scenarios (\emph{e.g.,} mental health support, education, and companionship), emotional intelligence~\cite{DBLP:conf/aaai/LubisSYN18,DBLP:conf/aaai/HosseiniC21,lian2025affectgpt,10.1609/aaai.v40i21.38832} has become essential for building user trust and sustaining long-term engagement. Consequently, empathetic response generation, which recognizes user emotions and generates understanding responses, has garnered significant research attention.
Early studies primarily focus on the \emph{text-to-text} paradigm, producing empathetic textual responses for textual user inputs~\cite{DBLP:conf/acl/LiuZDSLYJH20,DBLP:conf/acl/ZhouCWH23}.
Since users naturally express emotions through multimodal signals, and multimodal responses facilitate more vivid and effective empathetic support,  %  empathetic response generation
recent studies~\cite{zhang-etal-2024-stickerconv,fei-etal-2024-empathyear,DBLP:journals/corr/abs-2508-12854,DBLP:conf/www/ZhangM0HLC025,10.1145/3701716.3715200} have shifted toward a \emph{multimodal-to-multimodal} paradigm. These methods  exploit rich non-verbal emotional cues, such as speech prosody and facial expressions, enabling more accurate emotion perception and the delivery of  immersive empathetic support.
As a compelling instantiation of such immersive interaction, avatar-based multimodal empathetic response generation~\cite{DBLP:journals/corr/abs-2508-12854,DBLP:conf/www/ZhangM0HLC025,10.1145/3701716.3715200} has emerged as an  attractive research direction. As shown in Figure~\ref{Model_task}, it involves perceiving user emotions from multimodal inputs and synthesizing emotionally aligned text, expressive speech, and synchronized talking-face video via a specified digital avatar.

\begin{figure*}[!t]
    \centering
    \includegraphics[scale=0.64]{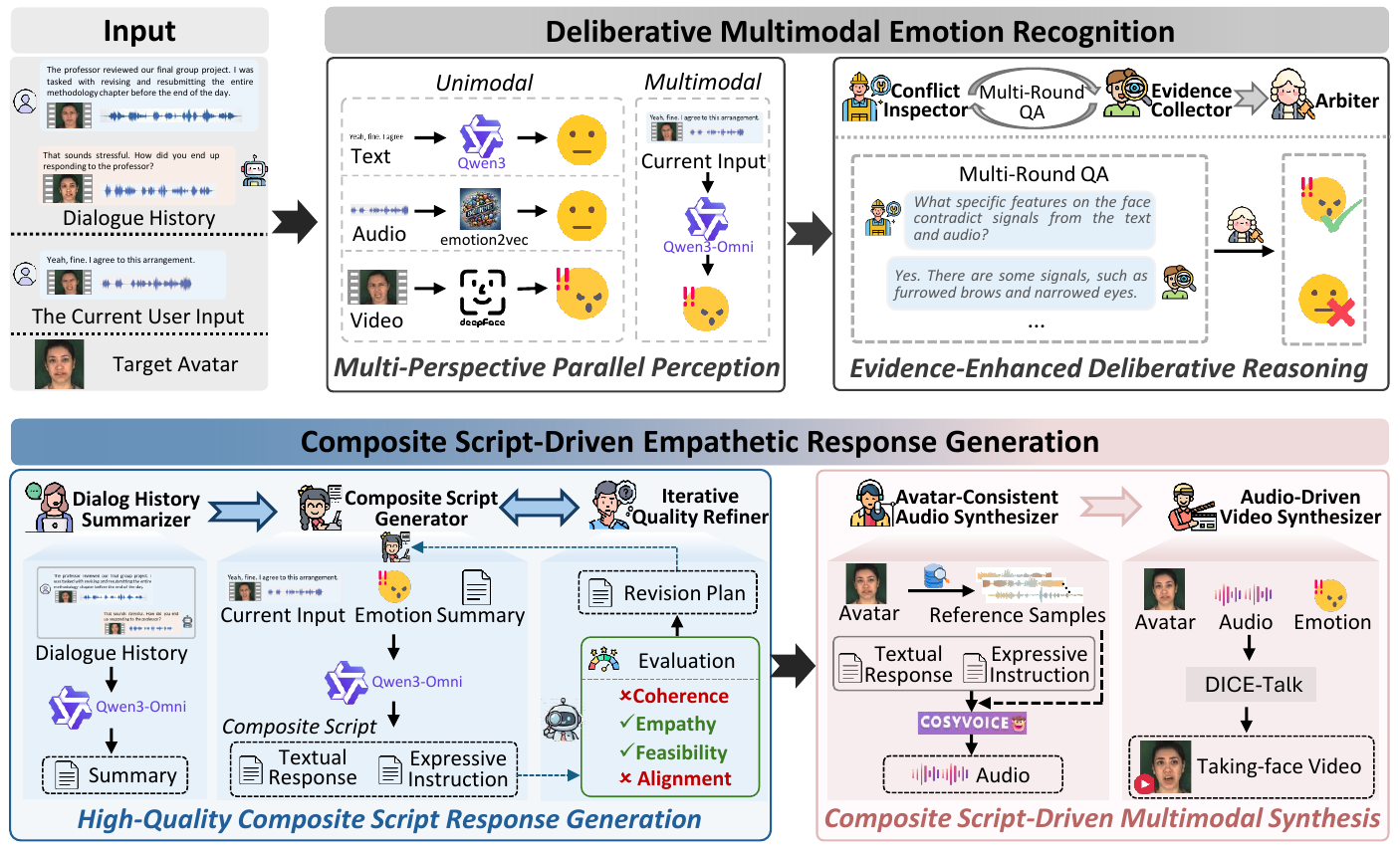}
    \caption{Illustration of the proposed framework. }
    % \vspace{-1em} , comprising two key modules: \textit{Deliberative Multimodal Emotion Recognition} and \textit{Composite Script-Driven Empathetic Response Generation}.
    \label{Model_figure}
\end{figure*}

Despite remarkable progress, existing methods remain constrained by the following limitations.
1) \textbf{Overlooking conflicting emotions across modalities}.
Existing studies typically feed all modalities into a Large Language Model (LLM) or a Multimodal LLM (MLLM) for emotion recognition, without examining inter-modal conflicts that often arise in scenarios such as sarcasm, emotional masking, or ambiguous expressions. 
For example, a user might write ``Great, another meeting'' in a superficially
positive tone, while their voice and facial expression clearly convey irritation.  
Without modeling such inter-modal conflicts, predictions can be incorrect. 
2) \textbf{Lacking expressive instructions for multimodal response generation}. 
Existing methods typically condition the audio and video generation on the textual response along with an implicit latent emotion vector or the recognized user emotion. While the textual response effectively conveys semantic content, it lacks explicit expressive instructions for guiding downstream audio and video synthesis. This can result in emotionally mismatched responses, such as producing a cheerful tone in a sad context or generating incongruent facial expressions in the talking-face video.
3) \textbf{Neglecting inherent error propagation in multimodal response generation}.
Existing studies follow a sequential pipeline (\emph{i.e.,} text $\rightarrow$ audio $\rightarrow$ video), where the textual response serves as the foundation for downstream modalities. Consequently, minor errors in the text stage can propagate, degrading the quality of the synthesized audio and video. However, existing studies lack mechanisms to ensure textual response quality, leaving this inherent error propagation unaddressed.

To address these limitations, we propose EmAvatar, which enhances multimodal
empathetic response generation via inter-modal conflict resolution and expressive
guidance, as shown in Figure~\ref{Model_figure}. It comprises two sequential tasks: emotion recognition and multimodal response generation.
For emotion recognition, the \textit{Deliberative Multimodal Emotion Recognition}
module first constructs an initial candidate set from independent unimodal and
multimodal predictions, explicitly exposing inter-modal conflicts. An
evidence-enhanced deliberative process then resolves them through multi-round
question-answering (QA) between a \textit{Conflict Inspector}, which poses
conflict-oriented questions, and an \textit{Evidence Collector}, which extracts
relevant evidence from the user input.
Finally, an \textit{Arbiter} consolidates the QA trajectory and candidate set into a reliable emotion prediction.
% , after which an \textit{Arbiter} consolidates
% the QA trajectory and the candidate set into a reliable emotion prediction. 
For response generation, the \textit{Composite Script-Driven Empathetic Response
Generation} module produces a composite script that couples the textual response with
an expressive instruction, specifying how the response should be realized in the audio
and video modalities based on a purified dialog history. An \textit{Iterative Quality
Refiner} further evaluates the script along multiple criteria and refines it according
to the identified deficiencies, so that the finalized script serves as explicit
guidance for downstream audio and video synthesis.
Extensive quantitative and qualitative experiments on four tasks demonstrate the effectiveness of our approach.

Our contributions are threefold.
\begin{itemize}
    \item We present a novel deliberative framework to address inter-modal emotional conflicts, a previously underexplored challenge within multimodal empathetic response generation.
    \item We pioneer the use of composite scripts to guide expressive response synthesis and introduce an iterative refinement mechanism to ensure script quality. 
    \item Extensive experiments across four tasks demonstrate the  superiority of EmAvatar. We release our code to facilitate future research.
\end{itemize} % \footnote{https://anonymous.4open.science/r/EmAvatar5659.}

\section{Related Work}
Early research on empathetic response generation focused on the text-to-text paradigm, only producing empathetic textual responses from user inputs~\cite{DBLP:conf/acl/LiuZDSLYJH20,DBLP:conf/acl/ZhouCWH23,DBLP:conf/aaai/LeNVNLNSLN26}. To capture critical non-verbal cues such as audio and video, several studies~\cite{wang-etal-2025-flexible,wu-etal-2025-traits,wu-etal-2025-personas,wu-etal-2024-ehdchat,Le2025ReinforcingTI,11129652,11146689,lian2025ovmer} incorporated multimodal information for improved emotion understanding. However,  outputs of these methods remain largely textual, overlooking multimodal responses (\emph{e.g.,} facial expressions and voice prosody) that are essential for vivid and deeply empathetic support.

In light of this, researchers have begun exploring {multimodal response generation}~\cite{zhang-etal-2024-stickerconv,DBLP:journals/corr/abs-2508-12854,10.1145/3701716.3715200,fei-etal-2024-empathyear}, especially the \mbox{avatar-based} empathetic generation to enhance emotional connection and facilitate face-to-face interactions. 
For example, Zhang et al.~\cite{DBLP:conf/www/ZhangM0HLC025} released a large-scale benchmark, AvaMERG, which supports generating multimodal responses spanning text, speech, and talking-face video. 
Additionally, Lin et al.~\cite{DBLP:journals/corr/abs-2508-12854} decomposed the task into empathy understanding, memory retrieval, and generation phases, leveraging the recognized user emotion to drive multimodal response synthesis.
Despite these advances, existing methods overlook inter-modal emotional conflicts, lack explicit guidance for response generation, and fail to ensure textual response quality, which is crucial for downstream multimodal generation.

\section{Methodology}

\subsection{Problem Formulation}
Suppose we have a set of $N$ dialog samples $\mathcal{D} = \{({\mathcal{H}_i}, {\mathcal{Q}_i}, {\mathcal{R}_i}, {e_i}, {a_i})\}_{i=1}^N$, where each sample contains the multimodal dialog history ${\mathcal{H}_i}$, the current multimodal user input  ${\mathcal{Q}_i}$,  the multimodal target response ${\mathcal{R}_i}$, the emotion label $e_i$ corresponding to  ${\mathcal{Q}_i}$, and the target avatar image  ${a_i}$ used to render ${\mathcal{R}_i}$.
For brevity, we  omit the subscript $i$ that indexes the samples.
Specifically, the dialog history is denoted as ${\mathcal{H}}=\{\mathcal{M}_j\}_{j=1}^{N_H}$, where $\mathcal{M}_j$ is the $j$-th multimodal utterance including the text, the speech audio, and the talking-face video, and $N_H$ is the number of utterances in ${\mathcal{H}}$.
The current user input is defined as ${\mathcal{Q}}=\{{\mathcal{T}_q}, {A_q}, {V_q}\}$, comprising the textual content ${\mathcal{T}_q}$, the speech audio ${A_q}$, and the talking-face video ${V_q}$.
Besides, the current input ${\mathcal{Q}}$ is associated with a ground-truth emotion label $e$ (\emph{e.g., } happy, angry, and surprised).
Similarly, the target response is represented as ${\mathcal{R}}=\{{\mathcal{T}_r}, A_r, V_r\}$, containing the text $\mathcal{T}_r$, speech audio $A_r$, and talking-face video $V_r$  rendered with the avatar image $a$. 
Based on the above dataset, we aim to obtain an MLLM-based training-free model $\mathcal{F}$ that can accurately recognize the user emotion and generate the multimodal empathetic response accordingly as follows, 
\begin{equation}
    \mathcal{F}({\mathcal{H}}, {\mathcal{Q}},  {a})\rightarrow{\{{e}, {\mathcal{R}}\}}.
     \label{eq1}
 \end{equation}

\subsection{Deliberative Multimodal Emotion Recognition}
Accurate emotion prediction is a prerequisite for generating precise empathetic
responses. To resolve inter-modal emotion conflicts, we introduce a Deliberative
Multimodal Emotion Recognition method with two stages: (1) \textit{Multi-Perspective
Parallel Perception}, which independently predicts candidate emotions from unimodal
and multimodal perspectives, thereby exposing potential \mbox{inter-modal} conflicts;
and (2) \textit{Evidence-Enhanced Deliberative Reasoning}, which probes these
conflicts by posing targeted discriminative questions and progressively gathering
evidence, producing an \mbox{evidence-aware} prediction.

% \subsection{Deliberative Multimodal Emotion Recognition}
% One prerequisite for generating the precise empathetic response is achieving  accurate user emotion prediction. To address the inter-modal emotion conflicts,  we introduce a Deliberative Multimodal Emotion Recognition method. It comprises two stages:
% (1) Multi-Perspective Parallel Perception, which independently predicts candidate emotion categories from diverse unimodal and multimodal perspectives, thereby laying the foundation for subsequent \mbox{inter-modal} conflict analysis.
% (2) Evidence-enhanced Deliberative Reasoning, which probes potential inter-modal conflicts by posing targeted discriminative questions and progressively gathering supporting evidence, ultimately producing an \mbox{evidence-aware} emotion prediction.

\subsubsection{Multi-Perspective Parallel Perception}

Specifically, we predict emotion categories from four perspectives: three unimodal views (\emph{i.e., } textual, acoustic, and visual) and one multimodal view.
For the textual content  ${\mathcal{T}_q}$, we adopt  Qwen$3$~\cite{yang2025qwen3technicalreport}  to identify the corresponding emotion $\hat{e}_t$ due to its superior capabilities in text semantic  understanding~\cite{gu2026escherverse}. 
Regarding the speech audio ${A_q}$, we classify the emotion $\hat{e}_a$ using the pretrained emotion2vec~\cite{DBLP:conf/acl/MaZYLGZ024}, which has demonstrated remarkable performance in speech emotion analysis~\cite{DBLP:conf/cvpr/ChenGHLTXWZZYWX25,DBLP:conf/nips/AoWTCZ0W0024}.
As for the video ${V_q}$, we utilize the pretrained  DeepFace~\cite{serengil2024lightface}  to identify the emotion $\hat{e}_v$, considering its excellent performance in the facial attribute analysis task~\cite{DBLP:conf/nips/WengYTQYL24}.
Additionally, to obtain a holistic understanding beyond isolated modalities, 
we further employ  Qwen3-Omni~\cite{Qwen3-Omni}, which has proven to be effective in multimodal emotional  understanding~\cite{DBLP:journals/corr/abs-2510-10444}.  Specifically, we feed all  three modalities (\emph{i.e.,} text, speech audio, and talking-face video) into Qwen3-Omni and recognize the overall emotion $\hat{e}_o$.
Consequently, we  acquire the initial emotion recognition set  $\mathcal{E}_p = \{{\hat{e}_t}, {\hat{e}_a}, {\hat{e}_v}, {\hat{e}_o}\}$.

\subsubsection{Evidence-Enhanced Deliberative Reasoning}
\label{sec:Evidence}

Ideally, if all predictions in $\mathcal{E}_p$ agree, we directly adopt $\hat{e}_o$ as
the final emotion. Otherwise, we perform evidence-enhanced deliberative reasoning, a
multi-round QA process in which a \textit{Conflict Inspector} raises targeted
discriminative questions on the inconsistencies within $\mathcal{E}_p$ (\emph{e.g.,}
``Does the facial expression suggest fear or anger?'' when $\hat{e}_a=$ ``fear''
conflicts with $\hat{e}_v=$ ``angry''), an \textit{Evidence Collector} answers them
with evidence extracted from the current user input, and an \textit{Arbiter}
consolidates the QA trajectory and the candidate set into the final prediction.
% Ideally, if all predictions in $\mathcal{E}_p$ are consistent, we can directly adopt the overall emotion ${\hat{e}_o}$ as the final emotion.
% Otherwise, we further perform evidence-enhanced deliberative reasoning,  a multi-round QA process between a {\textit{Conflict Inspector}} and an {\textit{Evidence Collector}} to progressively gather discriminative evidence for resolving inter-modal conflicts. Specifically, the {\textit{Conflict Inspector}} actively diagnoses inconsistencies within the candidate emotion set and raises targeted discriminative questions (e.g., ``Does the facial expression suggest fear or anger?'' when $\hat{e}_a=\text{    ``fear''}$ conflicts with $\hat{e}_v=\text{``angry''}$). The {\textit{Evidence Collector}} then answers these questions by extracting relevant evidence from the current user input. After multiple rounds of QA, an {\textit{Arbiter}} is employed to perform the final evidence-enhanced emotion prediction by jointly reviewing the entire QA trajectory and the initial candidate set to produce the final prediction.
Specifically, we employ Qwen3 as the {\textit{Conflict Inspector}} $\mathcal{C}_{ins}$, leveraging its superior text-based logical reasoning capabilities~\cite{gu2026escherverse} to formulate questions. At the $j$-th turn, it generates a discriminative question $\hat{q}_j$ conditioned on the initial prediction set ${\mathcal{E}_p}$ and the previous QA trajectory ${\hat{\mathcal{H}}_{j-1}}$, as follows:
\begin{equation}
    {\hat{q}_j} ={\mathcal{C}_{ins}}({\mathcal{E}_p}, {{\hat{\mathcal{H}}_{j-1}}}, P_{ins}), 
    \label{eq2_q}
\end{equation}
where ${\hat{\mathcal{H}}_{j-1}}= \{({\hat{q}_1}, {\hat{a}_1}), \cdots, ({\hat{q}_{j-1}}, {\hat{a}_{j-1}})\}$ denotes the QA trajectory of the previous $j-1$ turns, and ${\hat{\mathcal{H}}_{0}} = \emptyset$. $P_{ins}$ is the prompt for $\mathcal{C}_{ins}$. The {\textit{Evidence Collector}} $\mathcal{C}_{obs}$, implemented by Qwen3-Omni, then extracts evidence from the current multimodal input ${\mathcal{Q}}$ for answering ${\hat{q}_j}$ as follows:
\begin{equation}
    {\hat{a}_j} ={\mathcal{C}_{obs}}({\hat{q}_j}, {\mathcal{Q}}, P_{obs}), 
    \label{eq2_a}
\end{equation}
where ${\hat{a}_j}$ is the evidence collected to answer $\hat{q}_j$, and $P_{obs}$ is the prompt for $\mathcal{C}_{obs}$.

After $K$ rounds of QA interactions, the {\textit{Arbiter}} $\mathcal{C}_{arb}$ (also implemented by Qwen3-Omni for multimodal perception) integrates the initial predictions ${\mathcal{E}_p}$, the complete QA trajectory ${\hat{\mathcal{H}}_{K}}$, and the current user input  ${\mathcal{Q}}$ to obtain the final user emotion ${\bar{e}}$ as follows,
\begin{equation}
    {\bar{e}} ={\mathcal{C}_{arb}}({\mathcal{E}_p}, {\hat{\mathcal{H}}_{K}},  {\mathcal{Q}}, P_{arb}), 
    \label{eq2_d}
\end{equation}
where $P_{arb}$ is the corresponding prompt for $\mathcal{C}_{arb}$.

\subsection{Composite Script-Driven Empathetic Response Generation}
Conditioned on the recognized emotion, this module generates a content-synchronized
response across text, speech audio, and talking-face video. To explicitly guide
emotional expression in audio and video, it comprises two parts: 1)
\textit{High-Quality Composite Script Response Generation}, which produces a composite script
consisting of a textual response and an expressive instruction (\emph{e.g.,} ``speak
with a soothing, slow-paced tone'') that encodes how the response should be realized
in the audio and video modalities; and 2) \textit{Composite Script-Driven Multimodal
Synthesis}, which synthesizes expressive audio from the script and then generates the
avatar video conditioned on that audio.

% \subsection{Composite Script-Driven Empathetic Response Generation}
% This module works on generating a multimodal empathetic response across three content-synchronized modalities: text, speech audio, and talking-face video, based on the recognized user's emotion. 
% As a major novelty, to explicitly guide emotional expression in audio and video, we propose a Composite Script-Driven Empathetic Response Generation framework, comprising two modules:
% 1) High-Quality Composite Script Generation, which produces a composite script comprising a textual response and an expressive instruction (\emph{e.g.,} ``speak with a soothing, slow-paced tone'') that explicitly encodes how to realize the textual response in audio and video modalities. 
% 2) Composite Script-Driven Multimodal Synthesis, which uses the composite script to first synthesize expressive audio and then generate the avatar video conditioned on the synthesized audio. 

\subsubsection{High-Quality Composite Script Response Generation}
\label{sec:Comscript}

To ensure the  composite script quality,
we employ three specialized agents: \textit{Dialog History Summarizer} for purifying the context and extracting salient events; \textit{Composite Script Generator} for producing the initial textual response along with an expressive instruction, 
and \textit{Iterative Quality Refiner} for iteratively refining the script.

{\textbf{Dialog History Summarizer}.} 
Unlike approaches that feed the full raw history into response generation, we
introduce the \textit{Dialog History Summarizer} $\mathcal{A}_{sum}$, implemented with
Qwen3-Omni, to filter irrelevant content from the verbose multimodal history
$\mathcal{H}$ into a salient  context summary $M_{his}$:
% Unlike current approaches that use the full raw history for response generation, we introduce the \textit{Dialog History Summarizer} $\mathcal{A}_{sum}$ with Qwen3-Omni to filter irrelevant content from the verbose multimodal history $\mathcal{H}$, producing a concise semantic summary. Formally, the summarization is defined as:
\begin{equation}
    M_{his} = \mathcal{A}_{sum}(\mathcal{H}, P_{sum}),
\end{equation}
where $P_{sum}$ is the prompt for $\mathcal{A}_{sum}$.

{\textbf{Composite Script Generator}.} 
This agent is responsible for generating a composite script that includes both a textual response and an expressive instruction, based on the current input $\mathcal{Q}$, the predicted emotion $\bar{e}$, and the summary $M_{his}$ as: 
\begin{equation}
    (\bar{\mathcal{T}}_r, \mathcal{S}_{audio}) = \mathcal{A}_{com}(\mathcal{Q}, \bar{e}, M_{his}, P_{com}),
\end{equation}
where $\bar{\mathcal{T}}_r$ is the textual response, and $\mathcal{S}_{audio}$ is the expressive instruction that guides the expression of $\bar{\mathcal{T}}_r$ in the audio and video modalities. $P_{com}$ is the prompt for $\mathcal{A}_{com}$. Here, we employ Qwen3-Omni as $\mathcal{A}_{com}$.

{\textbf{Iterative Quality Refiner}.} 
To prevent errors from propagating to downstream audio and video synthesis, we propose
the \textit{Iterative Quality Refiner} $\mathcal{A}_{ref}$, which refines the composite
script until it satisfies four assessment dimensions: 1) \textbf{Coherence}, the
logical consistency between the textual response and the dialogue history; 2)
\textbf{Empathy}, the emotional appropriateness of the textual response; 3)
\textbf{Feasibility}, the practical executability of the expressive instruction; and
4) \textbf{Alignment}, the emotional consistency between the textual response and the
expressive instruction.

% To prevent errors from propagating to downstream audio and video synthesis, we propose the \textit{Iterative Quality Refiner} $\mathcal{A}_{ref}$, which iteratively refines the generated composite script until it satisfies four predefined assessment dimensions. Specifically, we jointly consider: 
% 1) \textbf{Coherence}, evaluating the logical consistency between the textual response and the dialogue history;
% 2) \textbf{Empathy}, measuring the emotional appropriateness of the textual response;
% 3) \textbf{Feasibility}, assessing the practical executability of the generated expressive instruction; 
% and 4) \textbf{Alignment}, evaluating the emotional consistency between the textual response and expressive instruction.

At the $k$-th iteration, $\mathcal{A}_{ref}$ evaluates the candidate script $(\bar{\mathcal{T}}_r^k, \mathcal{S}_{audio}^k)$ and produces a set of scores $\mathcal{R}_{ref}^k= \{s_1^k, s_2^k, s_3^k, s_4^k\}$. 
Each score $s_j^k$ is a discrete integer drawn from a 5-level scale (\emph{i.e.,} $s_j^k \in \{1, 2, 3, 4, 5\}$), quantifying the model's performance on the $j$-th assessment dimension from ``poor'' to ``excellent''.
%, where $s_j^k$ is the discrete integer evaluation score for the $j$-th assessment dimension. }
If any score in $\mathcal{R}_{ref}^{k}$ falls below a predefined threshold $\tau$ (\emph{i.e.,} $\exists s \in \mathcal{R}_{ref}^{k}, s < \tau$), $\mathcal{A}_{ref}$ then acts as a diagnostician, generating a natural language-based revision plan $\mathcal{P}_{rev}^{k}$ that aims to address  the identified deficiencies (\emph{e.g.,} ``The tone is overly cheerful; please soften the prosody''). The assessment process is formulated as:
\begin{equation}
(\mathcal{R}_{ref}^{k}, \mathcal{P}_{rev}^{k}) = \mathcal{A}_{ref}(\bar{\mathcal{T}}_r^{k}, \mathcal{S}_{audio}^{k},  M_{his}, \mathcal{Q}, \bar{e}, P_{ref}),
\end{equation}
where we use Qwen3-Omni as $\mathcal{A}_{ref}$ with prompt $P_{ref}$. 
The revision plan $\mathcal{P}_{rev}^{k}$ is then incorporated into the \textit{Composite Script Generator}  as additional guidance to refine the script in the next iteration:
\begin{equation}
(\bar{\mathcal{T}}_r^{k+1}, \mathcal{S}_{audio}^{k+1}) = \mathcal{A}_{com}(\mathcal{Q}, \bar{e}, M_{his}, \bar{\mathcal{T}}_r^k, \mathcal{S}_{audio}^k, \mathcal{P}_{rev}^{k}, P_{com}^{re}), 
\end{equation}
where $P_{com}^{re}$ is the prompt for the iterative refinement process. 
Upon meeting the pre-defined threshold, the iterative procedure terminates, producing the final refined textual response $\hat{\mathcal{T}}_r$ and emotional expressive instruction $\hat{\mathcal{S}}_{audio}$.

\subsubsection{Composite Script-Driven Multimodal Synthesis}
Equipped with the refined composite script, we employ two agents for the final
rendering: an \textit{Avatar-Consistent Audio Synthesizer} $\mathcal{A}_{syn}^a$ that
synthesizes speech preserving the target avatar's vocal identity, and an
\textit{Audio-Driven Video Synthesizer} $\mathcal{A}_{syn}^v$ that generates the
talking-face video aligned with the speech, the emotion, and the avatar.

% Equipped with the refined composite script, we employ two agents for the final
% rendering: an \textit{Avatar-Consistent Audio Synthesizer}  that
% synthesizes speech preserving the target avatar's vocal identity, and an
% \textit{Audio-Driven Video Synthesizer}  that generates the
% talking-face video aligned with the synthesized speech, the emotion, and the avatar.

% \subsubsection{Composite Script-Driven Multimodal Synthesis}
% \label{sec:multisys}
% Equipped with the refined composite script, we proceed to the final rendering stage. Here, we employ two specialized agents: an \textit{Avatar-Consistent Audio Synthesizer} to synthesize speech that faithfully maintains the target avatar's vocal identity  and an \textit{Audio-Driven Video Synthesizer} to generate the talking-face video aligning to the synthesized speech, the emotion, and the given avatar. 

\textbf{Avatar-Consistent Audio Synthesizer}.
Following previous studies~\cite{DBLP:conf/www/ZhangM0HLC025,DBLP:journals/corr/abs-2508-12854}, we design the \textit{Avatar-Consistent Audio Synthesizer} $\mathcal{A}_{syn}^a$, which first retrieves a set of acoustic reference samples $\mathcal{U}_{ref}$ associated with avatar $a$ from the dataset $\mathcal{D}$, and then synthesizes the audio $\bar{A}_r$ conditioned on the composite script, including the textual response $\hat{\mathcal{T}}_r$ and expressive instruction $\hat{\mathcal{S}}_{audio}$, as well as the retrieved references $\mathcal{U}_{ref}$:
\begin{equation}
        \bar{A}_r = \mathcal{A}_{syn}^a(\hat{\mathcal{T}}_r, \hat{\mathcal{S}}_{audio}, \mathcal{U}_{ref}).
    \label{eq_audio}
\end{equation}
We employ the pretrained CosyVoice2~\cite{du2024cosyvoice} for speech synthesis given its strength in text-to-speech tasks~\cite{chen-etal-2025-slam,hussain-etal-2025-koel}.

\textbf{Audio-Driven Video Synthesizer}.
Upon obtaining the synthesized audio,  we devise the \textit{Audio-Driven Video Synthesizer} $\mathcal{A}_{syn}^v$ that animates the static avatar image $a$ to produce the talking-face video $\bar{V}_r$.
Specifically, following~\cite{DBLP:journals/corr/abs-2508-12854}, we employ the pretrained DICE-Talk~\cite{tan2025dicetalk} for video generation. As an audio-driven model, DICE-Talk does not accept additional inputs; moreover, the expressive instruction is already embedded in the synthesized audio. Therefore, $\mathcal{A}_{syn}^v$ relies solely on the audio for video synthesis, with the predicted emotion $\bar{e}$ and specified avatar $a$ incorporated as additional constraints, as follows:
\begin{equation}
    \bar{V}_r = \mathcal{A}_{syn}^v(\bar{A}_r, \bar{e}, a).
\end{equation}

\section{Experiment}

\subsection{Experimental Setup}
%======================================================================

\textbf{Dataset}.
We evaluate EmAvatar on AvaMERG~\cite{DBLP:conf/www/ZhangM0HLC025}, currently the
only public benchmark providing synchronized text, speech, and talking-face video,
as required by all four of our tasks (text-only datasets such as
EmpatheticDialogues~\cite{rashkin-etal-2019-towards} support none of the multimodal
ones).
AvaMERG contains $33{,}048$ dialogues with $152{,}021$ utterances, each comprising
emotionally aligned text, speech audio, and talking-face video. It spans 10
real-world conversational topics and is annotated with seven emotion categories
(\emph{angry, contempt, disgusted, fear, happy, sad}, and \emph{surprised}), and
further provides 65 digital avatars covering four age groups and five races.

\begin{table}[t]
    \centering
    \small
    \caption{Performance on Emotion Recognition and Textual Response Generation.  $\uparrow$: higher is better. ``\textit{Improvement}'': relative improvement over the best baseline. }
    \label{tab:emotion_text}

    \resizebox{\linewidth}{!}{
        \setlength{\tabcolsep}{1.8mm}
        \begin{tabular}{l|ccc}
            \toprule
            Methods & Accuracy(\%) $\uparrow$ & Dist-$1$(\%) $\uparrow$ & Dist-$2$(\%) $\uparrow$ \\
            \midrule
            KEMP {\color{gray} (AAAI'22)} & $35.87$ & $0.41$ & $1.78$ \\
            CEM {\color{gray} (AAAI'22)} & $37.32$ & $0.50$ & $2.07$ \\
            CASE {\color{gray} (ACL'23)} & $40.96$ & $0.54$ & $2.14$ \\
            Empatheia {\color{gray} (WWW'25)} & $45.51$ & $2.73$ & $11.75$ \\
            Ola-7b {\color{gray} (ArXiv'25)} & $49.95$ & $3.02$ & $22.03$ \\
            DZ {\color{gray} (MM'25)} & $52.75$ & $3.06$ & $20.17$ \\
            R1-Omni {\color{gray} (ArXiv'25)} & $58.31$ & - & - \\
            Qwen3-Omni {\color{gray} (ArXiv'25)} & {$66.97$} & \underline{$4.07$} & \underline{$25.55$} \\
            It's MyGO!!!!! {\color{gray} (MM'25)} & \underline{$68.42$} & $3.93$ & $23.64$ \\
            \rowcolor{gray!15} \textbf{EmAvatar} & \textcolor{defblue}{\textbf{70.87}} & \textcolor{defblue}{\textbf{4.91}} & \textcolor{defblue}{\textbf{26.46}} \\
            \midrule
            \textit{Improvement} & \textit{+3.58\%} & \textit{+20.64\%} & \textit{+3.56\%} \\
            \bottomrule
        \end{tabular}
        \label{rq1_tab1}
    }
\end{table}

\noindent\textbf{Tasks and Metrics}.
We evaluate EmAvatar on four tasks: emotion recognition, textual response
generation, speech audio generation, and talking-face video generation.
For quantitative evaluation, we adopt Accuracy for emotion recognition, Distinct
metrics (Dist-1/2)~\cite{li-etal-2016-diversity} for textual generation, Mel
Cepstral Distortion (MCD)~\cite{407206} for speech generation, and
SSIM~\cite{995823}, Landmark Distance (F-LMD)~\cite{8953690}, and SyncNet scores
(Sync$_{cf}$, Sync$_{dis}$)~\cite{DBLP:conf/accv/ChungZ16a,ma2025playmate} for
talking-face video generation.
For human evaluation, following prior work~\cite{DBLP:conf/www/ZhangM0HLC025},
three annotators compare shuffled response pairs drawn from 50 random test
dialogues, and each pair is labeled as ``Win'', ``Loss'', or ``Tie''.
Textual responses are rated on Empathy, Coherence, Informativity, and Fluency.
Speech quality is measured by the 5-scale Mean Opinion Score
(MOS)~\cite{VISWANATHAN200555} for naturalness~\cite{ye2025emotional} and
Similarity MOS (SMOS)~\cite{lorenzotrueba18_odyssey} for speaker similarity.
Talking-face videos are assessed on the six dimensions listed in
Table~\ref{tab:human_eval_combined}.
LLM-based evaluation is provided in the Supplementary Material.

\begin{table}[t]
    \centering
    \small
    \caption{Performance on Speech Audio Generation.}
    \label{tab:speech_audio}

    \resizebox{\linewidth}{!}{
        \setlength{\tabcolsep}{4mm}
        \begin{tabular}{l|ccc}
            \toprule
            Methods & MCD $\downarrow$ & MOS $\uparrow$ & SMOS $\uparrow$ \\
            \midrule
            E3RG {\color{gray} (MM'25)} & $9.58$   & \underline{$4.55$} & $2.89$ \\
            CosyVoice3 {\color{gray} (ArXiv'25)} & $9.19$   & $4.41$ & $3.17$ \\
            VoxCPM {\color{gray} (ArXiv'25)} & $8.51$   & $2.30$ & $3.26$ \\
            CosyVoice2 {\color{gray} (ArXiv'25)} & \underline{$8.25$}   & $4.21$ & \underline{$3.33$} \\
            \rowcolor{gray!15} \textbf{EmAvatar} & \textcolor{defblue}{\textbf{8.23}} &   \textcolor{defblue}{\textbf{4.65}} & \textcolor{defblue}{\textbf{3.44}} \\
            \midrule
            \textit{Improvement} & \textit{+0.24\%} & \textit{+2.20\%} & \textit{+3.30\%} \\
            \bottomrule
        \end{tabular}
    }
\label{rq1_tab2}
\end{table}

\begin{table}[t]
    \centering
    \small
    \caption{Performance on Talking-face Video Generation.}
    \label{tab:video_gen}

    \resizebox{\linewidth}{!}{
        \setlength{\tabcolsep}{1.5mm}
        \begin{tabular}{l|cccc}
            \toprule
            Methods & SSIM $\uparrow$ & F-LMD $\downarrow$ & Sync$_{cf}$ $\uparrow$ & Sync$_{dis}$ $\downarrow$ \\
            \midrule
            E3RG  {\color{gray} (MM'25)} & $0.69$ & \underline{$2.15$} & $5.44$ & $2.35$ \\
            SadTalker {\color{gray} (CVPR'23)} & $0.75$ & $14.81$ & $2.65$ & $2.62$ \\
            Sonic  {\color{gray} (CVPR'25)} & $0.75$ & $2.21$ & \underline{$5.65$} & \underline{$2.17$} \\
            Empatheia {\color{gray} (WWW'25)} & \textcolor{defblue}{\textbf{0.77}} & $2.72$ & $2.76$ & - \\
            \rowcolor{gray!15} \textbf{EmAvatar} & \underline{$0.76$} & \textcolor{defblue}{\textbf{2.15}} & \textcolor{defblue}{\textbf{6.08}} & \textcolor{defblue}{\textbf{2.03}} \\
            \midrule
            \textit{Improvement} & - & \textit{0.00\%} & \textit{+7.61\%} & \textit{+6.45\%} \\
            \bottomrule
        \end{tabular}
    }
    \label{rq1_tab3}
\end{table}

\noindent\textbf{Baselines}.
We compare EmAvatar with three groups of competitive baselines.
1) \textit{Emotion Recognition and Textual Response Generation}: conventional
text-based methods (KEMP~\cite{Li_Li_Ren_Ren_Chen_2022},
CEM~\cite{DBLP:conf/aaai/SabourZH22}, and CASE~\cite{zhou-etal-2023-case}), the
LLM-based method Empatheia~\cite{DBLP:conf/www/ZhangM0HLC025}, MLLM-based methods
(Ola-7b~\cite{liu2025ola}, R1-Omni~\cite{Zhao2025R1OmniEO}, and
Qwen3-Omni~\cite{Qwen3-Omni}), and the champion (It's MyGO!!!!!) and runner-up
(DZ) solutions from the \textit{ACM Multimedia 2025 Grand
Challenge}\footnote{\url{https://avamerg.github.io/MM25-challenge.}}.
R1-Omni is designed specifically for omni-multimodal emotion recognition and is
therefore evaluated only on the emotion recognition task.
2) \textit{Speech Audio Generation}: the prior work
E3RG~\cite{DBLP:journals/corr/abs-2508-12854} and pretrained text-to-speech
methods (CosyVoice3~\cite{du2025cosyvoice}, VoxCPM~\cite{voxcpm2025}, and
CosyVoice2~\cite{du2024cosyvoice}).
3) \textit{Talking-face Video Generation}: existing empathetic methods
(Empatheia~\cite{DBLP:conf/www/ZhangM0HLC025} and
E3RG~\cite{DBLP:journals/corr/abs-2508-12854}) and pretrained talking-face
generation models (SadTalker~\cite{10204743} and Sonic~\cite{ji2025sonic}).

\noindent\textbf{Implementation Details}.
Both the number of QA rounds and the maximum refinement iterations are set to
$2$, and the score threshold $\tau$ is set to $4$.
All experiments are conducted on a server with 8 NVIDIA H200 GPUs.
% TODO: 目前正文只有 QA 轮数和 refinement 迭代次数的敏感性数据，
% 缺少 tau 的敏感性结果。建议补一组 tau 取 3/4/5 的实验，
% 否则这里的 "as validated by the sensitivity analyses" 对 tau 不成立。

%======================================================================

\subsection{Comparison with State-of-the-Art Methods}
\textbf{Quantitative Comparison}.
Tables~\ref{rq1_tab1}, \ref{rq1_tab2}, and \ref{rq1_tab3} report the results.
1) \textit{Emotion Recognition and Textual Response Generation}. In
Table~\ref{rq1_tab1}, EmAvatar performs best on all three metrics, surpassing
the challenge champion by 3.58\% in accuracy and the strongest baseline by
20.64\% in Dist-1, verifying that resolving inter-modal conflicts yields more
precise emotion prediction and thus more informative responses.
2) \textit{Speech Audio Generation}. In Table~\ref{rq1_tab2}, EmAvatar obtains
the best MOS and SMOS with a comparable MCD, indicating that the expressive
instruction improves naturalness and speaker similarity.
3) \textit{Talking-face Video Generation}. In Table~\ref{rq1_tab3}, EmAvatar
achieves the best lip synchronization and F-LMD with a comparable SSIM,
confirming its advantage on this most challenging task.
These consistent gains show that EmAvatar mitigates inter-modal conflicts,
provides explicit expressive guidance, and reduces error propagation in
sequential synthesis.

\noindent\textbf{Human Evaluation}.
We conduct a pairwise human evaluation against the strongest baselines,
Qwen3-Omni for textual responses and Sonic for multimodal responses.
As reported in Table~\ref{tab:human_eval_combined}, EmAvatar outperforms
Qwen3-Omni on all four textual dimensions. Against Sonic, it matches content
accuracy while winning clearly on empathy and multimodal consistency,
demonstrating more affectively appropriate and coherent multimodal outputs.

\subsection{Ablation Study}
%======================================================================

\begin{table}[!t]
    \centering
    \small
    \caption{Human evaluation results (\%) of EmAvatar versus Qwen3-Omni and Sonic on textual and multimodal dimensions. ``Consist.'' denotes Consistency.}
    \label{tab:human_eval_combined}
    \resizebox{\linewidth}{!}{\setlength{\tabcolsep}{1mm}
    \begin{tabular}{l|l|ccc}
        \toprule
        Comparison & Evaluation Factors & Win & Loss & Tie \\
        \midrule
        \multirow{4}{*}{\shortstack{EmAvatar vs.\\Qwen3-Omni\\(Textual)}}
        & Empathy        & 70.00 & 29.33 & 0.67 \\
        & Coherence      & 66.66 & 32.67 & 0.67 \\
        & Informativity  & 74.66 & 22.67 & 2.67 \\
        & Fluency        & 68.66 & 30.67 & 0.67 \\
        \midrule
        \multirow{6}{*}{\shortstack{EmAvatar vs.\\Sonic\\(Multimodal)}}
        & Speech Content Accuracy     & $24.00$ & 19.33 & 56.67 \\
        & Video Content Accuracy      & $24.00$ & 16.00 & 60.00 \\
        & Speech Empathy              & $68.66$ & 20.67 & 10.67 \\
        & Video Empathy               & $65.33$ & 26.00 & 8.67 \\
        & Multimodal Content Consist. & $50.00$ & 22.67 & 27.33 \\
        & Multimodal Emotional Consist.& $60.67$ & 30.00 & 9.33 \\
        \bottomrule
    \end{tabular}
    }
\end{table}

\begin{table}[t]
    \centering
    \small
    \caption{Ablation study results on emotion recognition.}
    \label{tab:ablation_emotion}

    % \resizebox{1\linewidth}{!}{
    \setlength{\tabcolsep}{12mm}

    \begin{tabular}{l|c}
        \toprule
        Methods & Accuracy (\%) $\uparrow$ \\ \midrule
        Video-Only & 45.08 \\
        Audio-Only & 46.56 \\
        Text-Only & 57.71 \\ \midrule
        w/o-Conflict & 66.97 \\
        Major-Vote & 66.20 \\
        Only-QA & 66.61 \\
        \rowcolor{gray!15} \textbf{EmAvatar} & \textcolor{defblue}{\textbf{70.87}} \\
        \bottomrule
    \end{tabular}
    % }
\end{table}

\begin{table}[t]
    \centering
    \small
    \caption{Generalization results on emotion recognition.}
    \label{tab:generalization_emotion}

    % \resizebox{1\linewidth}{!}{
    \setlength{\tabcolsep}{8mm}
    \begin{tabular}{l|cc}
        \toprule
        Backbones & Alone & w-Ours \\ \midrule
        Ola-$7$b & 49.95 & 62.93 \\
        R1-Omni & 58.31 & 68.76 \\ \midrule
        Qwen2-VL & 57.55 & {59.08} \\
        Qwen2.5-VL & 62.54 & {67.16} \\
        \rowcolor{gray!15} Qwen3-Omni & 66.97 & \textcolor{defblue}{\textbf{70.87}} \\
        \bottomrule
    \end{tabular}
    % }
\end{table}

\begin{table}[t]
    \centering
    \small
    \caption{Ablation results on the expressive instruction and its transfer to other text-to-speech models.}
    \label{tab:ablation_speech}

    \resizebox{\linewidth}{!}{
        \setlength{\tabcolsep}{3mm}
        \begin{tabular}{l|ccc}
            \toprule
            Methods & MCD $\downarrow$ & MOS $\uparrow$ & SMOS $\uparrow$ \\
            \midrule
            VoxCPM & $8.51$ & $2.30$ & $3.26$ \\
            VoxCPM-w-Instruction & $8.45$ & $2.34$ & $3.35$ \\ \midrule
            CosyVoice3 & $9.19$ & $4.41$ & $3.17$ \\
            CosyVoice3-w-Instruction & $8.41$ & $4.59$ & $3.25$ \\ \midrule
            w/o-Instruction & $8.25$ & $4.21$ & $3.33$ \\
            \rowcolor{gray!15} \textbf{EmAvatar} & \textcolor{defblue}{\textbf{8.23}} & \textcolor{defblue}{\textbf{4.65}} & \textcolor{defblue}{\textbf{3.44}} \\
            \bottomrule
        \end{tabular}
    }
\end{table}

\begin{table}[t]
    \centering
    \small
    \caption{Sensitivity results on the number of script refinement iterations. Iter.~1 corresponds to the w/o-Refiner variant.}
    \label{tab:refine_sensitivity}

    \resizebox{1\linewidth}{!}{
        \setlength{\tabcolsep}{6mm}
        \begin{tabular}{l|cc}
            \toprule
            {Methods} & Dist-1(\%) $\uparrow$ & Dist-2(\%) $\uparrow$ \\
            \midrule
            Iter.~1 (w/o-Refiner) & $4.85$ & $24.86$ \\
            \rowcolor{gray!15} \textbf{Iter.~2 (EmAvatar)} & \textcolor{defblue}{\textbf{4.91}} & \textcolor{defblue}{{26.46}} \\
            Iter.~3 & $4.89$ & $26.32$ \\
            Iter.~4 & $4.87$ & {26.73} \\
            \bottomrule
        \end{tabular}
    }
\end{table}

\textbf{On Emotion Recognition}.
To verify the necessity of conflict handling and locate the source of the gains,
we design three variants: (1) w/o-Conflict,  predicting emotion directly with
Qwen3-Omni; (2) Major-Vote,  taking a majority vote over the four initial
predictions without QA reasoning; and (3) Only-QA,  performing multi-round QA
on the user input without the specialized model priors. We also report
Text-Only, Audio-Only, and Video-Only, where prediction relies on a single
modality.
Table~\ref{tab:ablation_emotion} yields three observations. First, EmAvatar
outperforms w/o-Conflict, Major-Vote, and Only-QA, indicating  the gain
comes from genuine conflict resolution rather than the backbone itself, simple
prediction aggregation, or extra LLM inference. Second, w/o-Conflict still
surpasses the three single-modality variants, confirming  different
modalities carry complementary cues. Third, Text-Only outperforms Audio-Only and
Video-Only, showing  text dominates emotion recognition owing to its
explicit semantics.

\noindent\textbf{On Multimodal Response Generation}.
We examine two variants: w/o-Refiner, which generates the composite script in a
single pass, and w/o-Instruction, which omits the prosodic guidance in
Eqn.~(\ref{eq_audio}).
As shown in Table~\ref{tab:refine_sensitivity} (Iter.~1) and
Table~\ref{tab:ablation_speech}, EmAvatar outperforms w/o-Refiner on textual
generation and w/o-Instruction on speech synthesis, confirming the value of
iterative refinement and explicit prosodic guidance.
% TODO: 建议给 w/o-Refiner 补充下游音频与视频指标 (MCD/MOS/SSIM/Sync)，
% 以直接支撑引言中的第三个 limitation (误差传播)。目前 Dist-1/2
% 只能反映文本多样性，无法说明 refiner 阻断了误差向下游传播。

%======================================================================
\subsection{Generalization Study}
%======================================================================

We examine generalization from two aspects.
\textbf{Backbone Generalization}. We plug our deliberative framework into the
two strongest public baselines (Ola-$7$b and R1-Omni) and into
earlier-generation backbones (Qwen2-VL/Qwen2 and Qwen2.5-VL/Qwen2.5).
Table~\ref{tab:generalization_emotion} shows consistent gains on all five
backbones, verifying that the improvement stems from the framework design
rather than the latest backbone.
\textbf{Instruction Transfer}. We further apply the expressive instruction to
VoxCPM and CosyVoice3. As shown in Table~\ref{tab:ablation_speech}, both
instruction-enhanced variants surpass their backbones on all metrics,
confirming that the instruction transfers well across speech models.
% We examine generalization from two aspects.
% \textbf{Backbone Generalization}. We plug our deliberative framework into the
% two strongest publicly available baselines (Ola-$7$b and R1-Omni) and into
% earlier-generation backbones (Qwen2-VL/Qwen2 and Qwen2.5-VL/Qwen2.5).
% Table~\ref{tab:generalization_emotion} shows that our framework improves every
% backbone, including Ola-$7$b, R1-Omni, and three generations of Qwen models,
% verifying that the gains stem from the framework design rather than the latest
% backbone.
% \textbf{Instruction Transfer}. We further transfer the expressive instruction to
% two other text-to-speech models, yielding VoxCPM-w-Instruction and
% CosyVoice3-w-Instruction. As shown in Table~\ref{tab:ablation_speech}, both
% instruction-enhanced variants improve over their backbones across all metrics,
% showing that the expressive instruction transfers well across text-to-speech
% models.

%======================================================================
\subsection{Sensitivity Analysis}
\label{sec:sensitivity}
%======================================================================

We analyze the two key hyperparameters of EmAvatar.
\textbf{Number of QA Rounds}. The accuracy with 1, 2, and 3 rounds is 70.75\%,
70.87\%, and 70.78\%, respectively, showing that performance is stable and 2
rounds are sufficient.
\textbf{Number of Refinement Iterations}. We vary the maximum refinement
iterations from 1 to 4. As shown in Table~\ref{tab:refine_sensitivity},
increasing the iterations beyond 2 brings no consistent gain, so 2 iterations
strike a favorable balance between quality and cost.
% TODO: 建议将 QA 轮数、refinement 迭代次数、tau 合并为一张三联子图，
% 既节省版面，也能补上 tau 的敏感性缺口。

%======================================================================
\subsection{Inference Efficiency Analysis}
%======================================================================

\begin{table}[!t]
    \centering
    \small
    \caption{Comparison of inference efficiency (s). Baseline models are
    Qwen3-Omni for reasoning and script generation, CosyVoice2 for audio, and Sonic
    for video.}
    \label{tab:latency_final_v3}
    \setlength{\tabcolsep}{2.1mm}
    \begin{tabular}{l|l|cc}
        \toprule
        Tasks & Modules & EmAvatar & Baseline \\
        \midrule
        \multirow{4}{*}{\shortstack[l]{Emotion\\ Recog.}}
            & Conflict Inspector & 2.89 & \multirow{3}{*}{1.69} \\
            & Evidence Collector & 8.03 & \\
            & Arbiter & 1.44 & \\
            \cmidrule{2-4}
            & Subtotal & 12.36 & 1.69 \\
        \midrule
        \multirow{6}{*}{\shortstack[l]{Multimodal\\ Response\\ Generation}}
            & Dialog Summarizer & 8.10 & \multirow{3}{*}{5.05} \\
            & Script Generator & 17.10 & \\
            & Quality Refiner & 4.10 & \\
            \cmidrule{2-4}
            & Audio Synthesizer & 3.46 & 1.92 \\
            & Video Synthesizer & 61.81 & 88.35 \\
            \cmidrule{2-4}
            & Subtotal & 94.57 & 95.32 \\
        \midrule
        \multicolumn{2}{l|}{\textbf{Total Cost}} & \textcolor{defblue}{\textbf{106.93}} & \textbf{97.01} \\
        \bottomrule
    \end{tabular}
\end{table}

Table~\ref{tab:latency_final_v3} reports the inference cost of EmAvatar and the
baseline. EmAvatar completes the pipeline in 106.93 seconds, only 9.92 seconds
above the baseline, so the deliberative and iterative design adds limited
overhead. Emotion recognition takes 12.36 seconds, of which the Evidence
Collector accounts for 8.03 seconds. % due to evidence extraction from multimodal
input.
This deliberative conflict-resolution process brings a 3.90-point accuracy gain over the single-pass baseline, a favorable trade-off for conflict-aware recognition.
% which brings a 3.90-point accuracy gain over the single-pass baseline, a
% favorable trade-off for conflict-aware recognition. 
Multimodal response
generation takes 94.57 seconds, slightly below the 95.32-second baseline, and is
dominated by the Video Synthesizer at 61.81 seconds, while the Dialog
Summarizer, Script Generator, and Quality Refiner together take only 29.30
seconds.

%======================================================================
\subsection{Case Study}
%======================================================================

\begin{figure}[!t]
    \centering
    \includegraphics[scale=0.37]{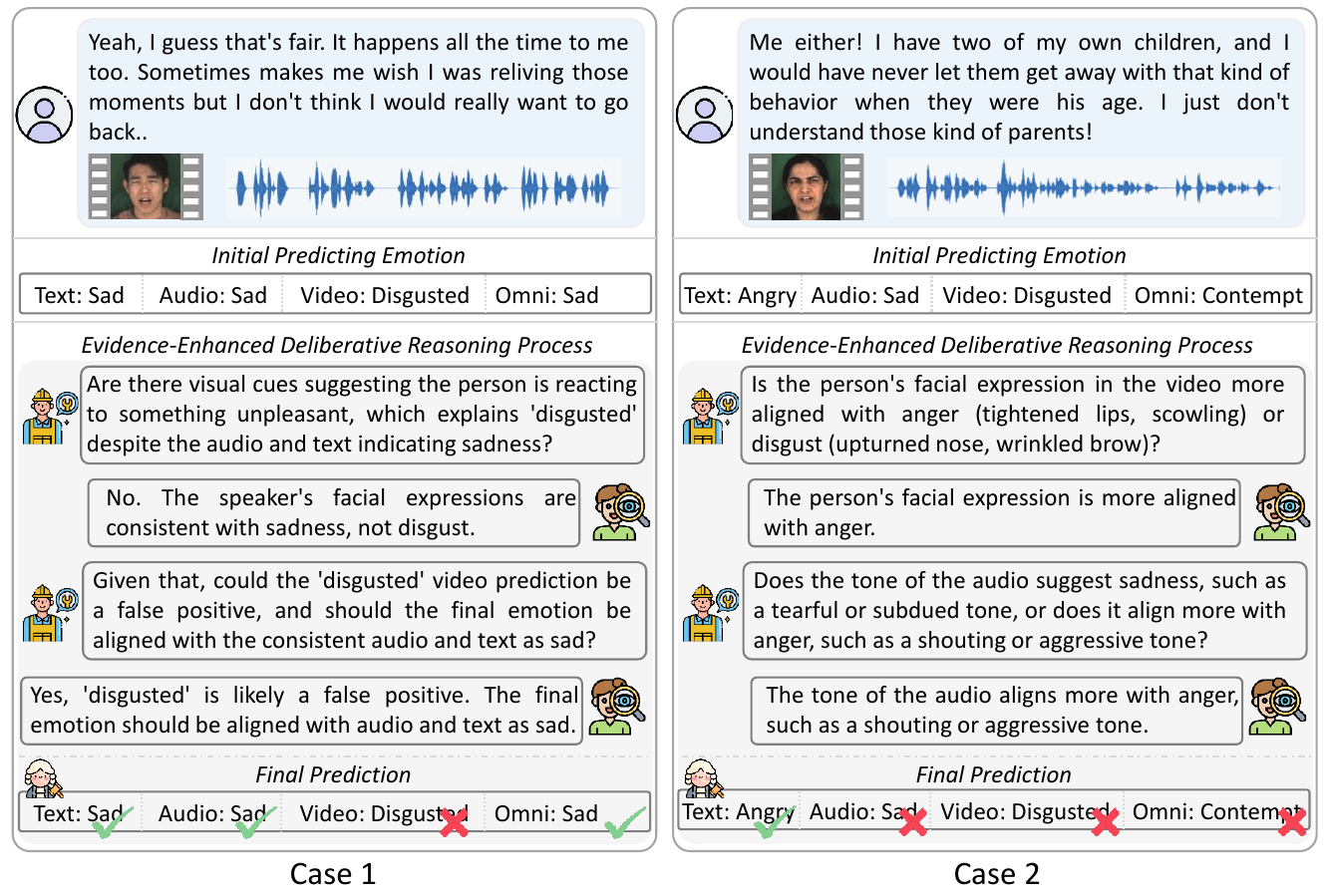}
    \caption{Two test cases showing how deliberative multimodal recognition resolves inter-modal conflicts.}
    \label{rq2_figure}
\end{figure}

\begin{figure}[!t]
    \centering
    \includegraphics[scale=0.37]{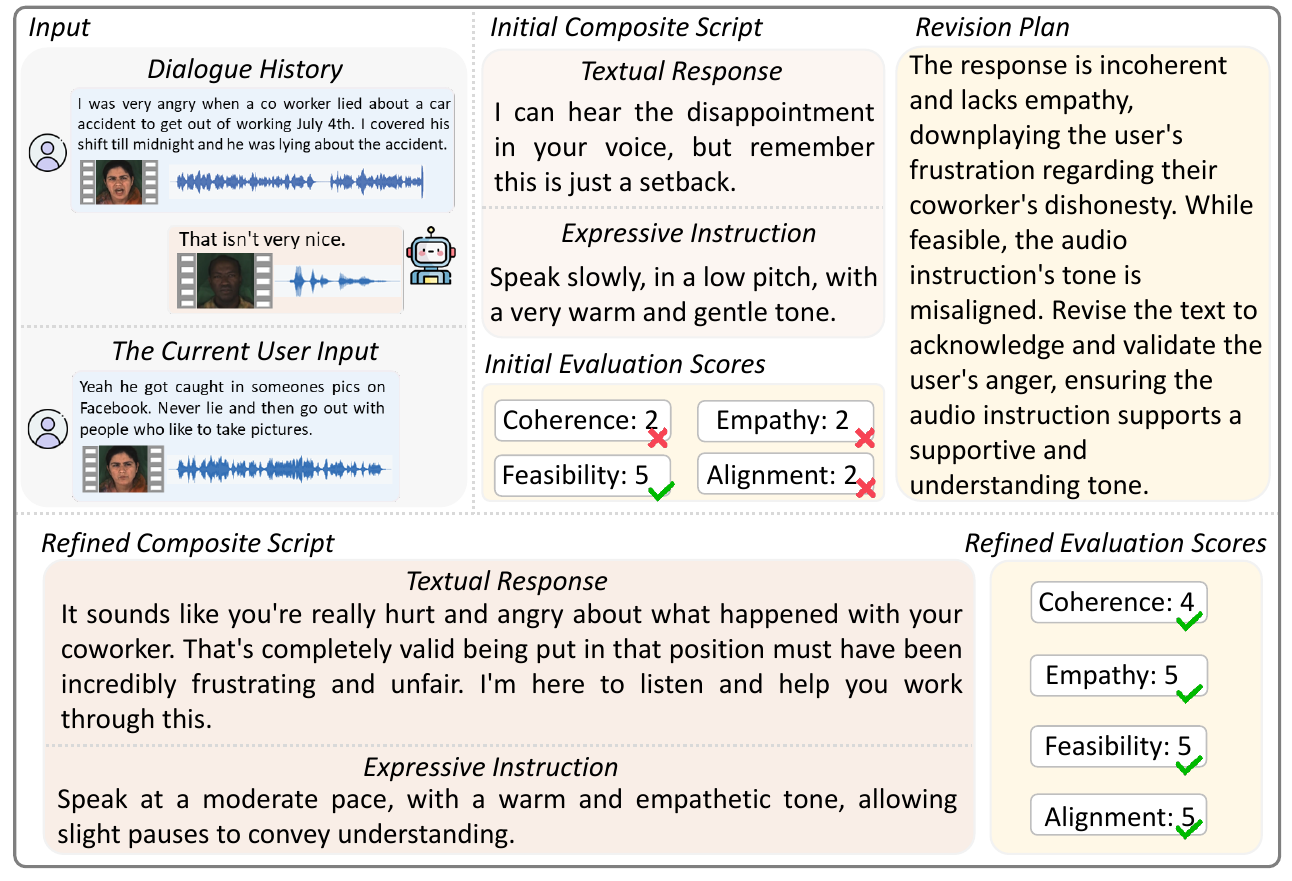}
    \caption{Illustration of iterative composite script refinement. Ticks and crosses denote satisfied and unsatisfied criteria.}
    \label{rq3_figure}
\end{figure}

\textbf{On Conflict Resolution}.
Figure~\ref{rq2_figure} illustrates how deliberative reasoning resolves
inter-modal conflicts. In \textit{Case 1}, the user recalls the past with
regret; the text and audio branches correctly predict ``Sad'', whereas the video
branch predicts ``Disgusted''. The \textit{Conflict Inspector} diagnoses this
inconsistency and raises a targeted question, and the \textit{Evidence
Collector} re-examines the video and finds the facial cues consistent with
sadness. The \textit{Arbiter} therefore rejects ``Disgusted'' and settles on
``Sad''. \textit{Case 2} follows the same pattern, confirming that inter-modal
conflicts do occur and that our deliberative reasoning corrects them.

\noindent\textbf{On Script Refinement}.
Figure~\ref{rq3_figure} visualizes the iterative refinement process of one
testing sample. The user expresses strong anger at a coworker's dishonesty, yet
the initial script dismisses this as ``just a setback'' and prescribes a generic
``gentle'' tone, receiving low scores on Coherence, Empathy, and Alignment. The
Revision Plan pinpoints this misalignment, and the refined script validates the
user's feelings (\emph{e.g.,} ``that's completely valid'') and switches to a
``moderate pace'' with an ``empathetic tone'', satisfying all four criteria.
This shows that iterative refinement reliably improves script quality and that
the expressive instruction provides explicit directives for downstream
synthesis. More cases are given in the Supplementary Material.
\section{Conclusion}
In this work, we propose EmAvatar, a novel model for multimodal empathetic response generation, consisting of two modules: \textit{Deliberative Multimodal Emotion Recognition} for resolving inter-modal conflicts and \textit{Composite Script-Driven Empathetic Response Generation} for generating multimodal responses with explicit expressive guidance.
Quantitative and qualitative experiments on four tasks comprehensively verify the effectiveness of EmAvatar. 
Case studies further confirm the existence of cross-modal emotion conflicts and show that EmAvatar effectively mitigates them through deliberative reasoning. 
In addition, experiments demonstrate that iterative quality refinement is crucial for ensuring the reliability of composite scripts and for providing precise guidance in expressive multimodal response synthesis.
In the  future, we plan to enhance the  inference efficiency to better support real-time applications and practical deployment. % support practical deployment. %  we plan to further optimize the computational efficiency of EmAvatar to facilitate more real-time and seamless human-avatar interactions.

\clearpage
\bibliography{sample-base}

\clearpage
% \appendix
% \input{Appendix}

\end{document}